\documentclass[conference]{IEEEtran}
\IEEEoverridecommandlockouts
\usepackage{cite}
\usepackage{amsmath,amssymb,amsfonts}
\usepackage{algorithmic}
\usepackage{graphicx}
\usepackage{textcomp}
\usepackage{xcolor}
\usepackage{hyperref}
\def\BibTeX{{\rm B\kern-.05em{\sc i\kern-.025em b}\kern-.08em
    T\kern-.1667em\lower.7ex\hbox{E}\kern-.125emX}}
\begin{document}

%
%

\title{Verif.ai: Towards an Open-Source Scientific Generative Question-Answering System with Referenced and Verifiable Answers\\
\thanks{NGI Search}
}
\author{\IEEEauthorblockN{1\textsuperscript{st} Miloš Košprdić}
\IEEEauthorblockA{\textit{Institute for Artificial Intelligence Research} \\
\textit{ and Development of Serbia}\\
Fruškogorska 1, Novi Sad, Serbia \\
email: milos.kosprdic@ivi.ac.rs}
\and
\IEEEauthorblockN{2\textsuperscript{nd} Adela Ljajić}
\IEEEauthorblockA{\textit{Institute for Artificial Intelligence Research} \\
\textit{ and Development of Serbia}\\
Fruškogorska 1, Novi Sad, Serbia \\
email: adela.ljajic@ivi.ac.rs}
\and
\IEEEauthorblockN{3\textsuperscript{rd} Bojana Bašaragin}
\IEEEauthorblockA{\textit{Institute for Artificial Intelligence Research} \\
\textit{ and Development of Serbia}\\
Fruškogorska 1, Novi Sad, Serbia \\
email: bojana.basaragin@ivi.ac.rs}
\and
\IEEEauthorblockN{4\textsuperscript{th} Darija Medvecki}
\IEEEauthorblockA{\textit{Institute for Artificial Intelligence Research} \\
\textit{ and Development of Serbia}\\
Fruškogorska 1, Novi Sad, Serbia \\
email: darija.medvecki@ivi.ac.rs}
\and
\IEEEauthorblockN{5\textsuperscript{th} Nikola Milošević}
\IEEEauthorblockA{\textit{R\&D Data Sciences and AI} \\
\textit{Bayer A.G.}\\
Müllerstraße 178, Berlin, Germany \\
email: nikola.milosevic@bayer.com}
}

\maketitle

\begin{abstract}
In this paper, we present the current progress of the project \textbf{Verif.ai}, an open-source scientific generative question-answering system with referenced and verified answers. The components of the system are (1) an information retrieval system combining semantic and lexical search techniques over scientific papers (PubMed), (2) a fine-tuned 
generative model (Mistral 7B) taking top answers and generating answers with references to the papers from which the claim was derived, and (3) a verification engine that cross-checks the generated claim and the abstract or paper from which the claim was derived, verifying whether there may have been any hallucinations in generating the claim. We are reinforcing the generative model by providing the abstract in context, but in addition, an independent set of methods and models are verifying the answer and checking for hallucinations. Therefore, we believe that by using our method, we can make scientists more productive, while building trust in the use of generative language models in scientific environments, where hallucinations and misinformation cannot be tolerated.
\end{abstract}


\section{Introduction}
In recent years, the advent of large language models has revolutionized various domains, offering unprecedented capabilities in natural language understanding, generation, and interaction \cite{openai2023gpt4,jiang2023mistral,bubeck2023sparks,park2023generative,touvron2023llama,katz2023gpt}. Particularly within the scientific community, these models hold tremendous potential for accelerating research processes, automating information retrieval \cite{lewis2020retrieval}, and enhancing the generation of complex scientific content. However, as these models become integral to scientific workflows, a critical challenge emerges – the issue of hallucinations, or the inadvertent generation of false or misleading information \cite{wadden2020fact,huang2023survey,malaviya2023expertqa}.

In scientific domains where accuracy and reliability are paramount, the occurrence of hallucinations poses a significant impediment to the widespread adoption of large language models (LLMs) \cite{boyko2023interdisciplinary}. The potential for misinformation introduces an inherent trust deficit, hindering scientists from fully embracing generative language models. It is imperative to address this challenge comprehensively to ensure that the benefits of these models are harnessed without compromising the integrity of scientific knowledge.

In response to this pressing concern, we introduce the \textbf{Verif.ai} project, an open-source initiative aimed at mitigating the risk of hallucinations in scientific generative question-answering systems. Our approach relies on information retrieval, leveraging both semantic and lexical techniques over a vast repository of scientific papers such as PubMed\footnote{\url{https://pubmed.ncbi.nlm.nih.gov/}}, complemented with retrieval-augmented generation (RAG) using a fine-tuned generative model, Mistral 7B, for answer generation with traceable references. Notably, the system goes beyond mere answer generation by incorporating a verification engine that cross-checks the generated claims against the abstracts or papers from which they are derived. We believe that the system, which makes the best effort to indicate possible hallucinations to the user, coupled with hallucination reduction techniques, its open-source nature, and community support, will instill trust in the scientific community in the use of LLM-based scientific systems.

\section{Methodology}

\begin{figure}[!h]
     \centering
     \includegraphics[width=1.0\linewidth]{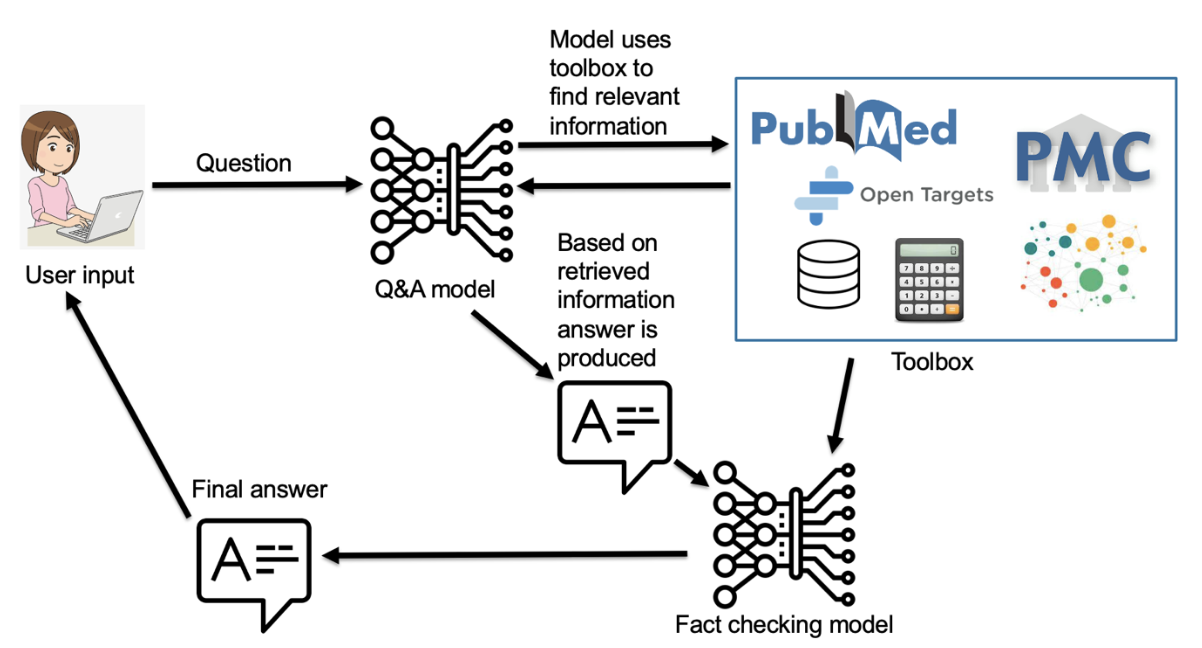}
     \caption{Methodology overview of the \textbf{Verif.ai} project}\label{Fig:InputTransformation}
\end{figure}

Our methodology employs a toolbox to discover relevant information and provide context to the question-answering system. Currently, the primary component of this toolbox is the information retrieval engine (PubMed). The question-answering system utilizes a fine-tuned LLM to generate answers based on the information from the toolbox. A fact-checking or verification engine examines the generated answer within the toolbox, identifying any potential hallucinations in the system. The final component of the system is a user interface, enabling users to ask a question, review answers and offer a feedback functionality, so they can contribute to the improvement of the \textbf{Verif.ai} project. The overview of the methodology is depicted in Figure \ref{Fig:InputTransformation}. In the following subsections, we provide details of the methods envisioned for each of the components.

\subsection{Toolbox and Information Retrieval}

The major component that has been implemented so far in our toolbox is the information retrieval engine. Our information retrieval engine is based on OpenSearch\footnote{\url{https://opensearch.org/}}, an open-source engine that was forked from Elasticsearch and is under the Apache 2 license. We have indexed PubMed articles using lexical indexing provided by OpenSearch. Additionally, we have created an index storing embeddings of documents using the MSMARCO model for semantic search. This model was selected because it can handle asymmetric searches (e.g., different lengths of queries compared to the searched texts) \cite{craswell2021ms}. Embeddings were stored in the OpenSearch field, allowing for the combination of lexical and semantic search. This approach emphasizes direct matches while also finding semantically similar phrases and parts of the text where the text does not match. The user question is first transformed into a query, and the most relevant documents are retrieved before being passed to the LLM that generates the answer.

\subsection{Question-answering with references}

For generating the answers, we have used Mistral 7B parameter model with instruction fine-tuning\footnote{\url{https://huggingface.co/filipealmeida/Mistral-7B-Instruct-v0.1-sharded}}. This model was further fine-tuned using questions from PubMedQA dataset \cite{jin2019pubmedqa} and generated answers using GPT3.5 with the most relevant documents from PubMed passed as context. The following prompt was used to generate answers:

\noindent\fbox{%
    \parbox{\linewidth}{%
Please carefully read the question and use the provided research papers to support your answers. When making a statement, indicate the corresponding abstract number in square brackets (e.g., [1][2]). Note that some abstracts may appear to be strictly related to the instructions, while others may not be relevant at all.
    }
}
\\

We have selected 10,000 random PubMedQA questions to generate this dataset. The dataset was then used to fine-tune the Mistral 7B model using the QLoRA methodology \cite{dettmers2023qlora}. The training was performed using a rescaled loss, a rank of 64, an alpha of 16, and a LoRA dropout of 0.1, resulting in 27,262,976 trainable parameters. The input to the training contained the question, retrieved documents (as many as can fit into the context), and the answer. We made this preliminary generated QLoRA adapter available on Hugging Face\footnote{\url{https://huggingface.co/BojanaBas/Mistral-7B-Instruct-v0.1-pqa}}. 

We then used the fine-tuned model for answer generation. Using the exactly same input as in training did not produce the expected results, and therefore, we added an instruction at the beginning of the prompt:

\noindent
\qquad \\\fbox{
    \parbox{\linewidth}{
       [INST] Answer the question using the given abstracts. Reference claims with the relevant abstract id in brackets (e.g. (PUBMED:123456) at the end of the sentence). Answer may contain references to many abstracts. Be as factual as possible and always use references in brackets. Use exclusively provided abstracts and their ids.
Make answer look similar to the following: Several genes play role in breast cancer. For example BRAC1, BRAC2 are well studied targets (PUBMED:554433). The other targets involve IRAK4, CAS2 and HMPA (PUBMED:665544).
    }
}
\\

The instruction was followed by the set of relevant documents obtained by querying OpenSearch and the question asked by the user. To prompt Mistral-7B-Instruct-v0.1-pqa, we use the mentioned template and default parameters with only two differences: we set max\_new\_tokens to 1000 and repetition\_penalty to 1.1.

\subsection{Verifying claims from the generated answer}

The aim of the verification engine is to parse sentences and references from the answer generation engine and verify that there are no hallucinations in the answer. Our assumption is that each statement is supported by one or more references. For verification, we compare the XLM-RoBERTa-large model\footnote{\url{https://huggingface.co/xlm-roberta-large}} and DeBERTa model\footnote{\url{microsoft/deberta-v3-large}}, treating it as a natural language inference problem. The selected model has a significantly different architecture than the generation model and is fine-tuned using the SciFact dataset \cite{wadden2020fact}. The dataset is additionally cleaned (e.g., claims were deduplicated, and instances with multiple citations in no-evidence examples were split into multiple samples, one for each reference). The input to the model contains the CLS token, the statement, a separator token, and the joined referenced article title and abstract, followed by another separation token. The output of the model falls into one of three classes:
\begin{itemize}
    \item \textbf{Supports} - in case statement is supported by the content of the article
    \item \textbf{Contradicts} - in case the statement contradicts the article
    \item \textbf{No Evidence} - in case there is no evidence in the article for the given claim
\end{itemize}
The fine-tuned model serves as the primary method for flagging contradictions or unsupported claims. However, additional methods for establishing user trust in the system will be implemented, including presenting to the user the sentences from the abstracts that are most similar to the claim.

\subsection{User feedback integration}

The envisioned user interface would present the answer to the user's query, referencing documents containing the answer and flagging sentences that contain potential hallucinations. However, users are asked to critically evaluate answers, and they can provide feedback either by changing a class of the natural language inference model or even by modifying generated answers. These modifications are recorded and used in future model fine-tuning, thereby improving the system.

\section{Preliminary evaluation}

In this section, we present the results based on our preliminary evaluation. At the time of writing of this article, the project was in the 3rd month of implementation, and we are working on improving our methodology and creating a web application that integrates all the described components.

\subsection{Information retrieval}

We have qualitatively evaluated OpenSearch's results on a small set of indexed PubMed articles. We compared lexical search, semantic search, and a hybrid combination of both lexical and semantic search. We observed that lexical search may perform better when the search terms can be exactly matched in the documents, while semantic search works well with paraphrased text or synonymous terms. Hybrid search managed to find documents containing terms that could be exactly matched, as well as ones that were paraphrased or contained synonyms. While semantic search would also find documents that contained an exact match of the terms, it often happened that they were not prioritized. Hybrid search helped in putting such documents at the top of the search results. Based on several user discussions, we have concluded that users expect the top results to be based on exact matches and later to find relevant documents that do not contain the searched terms.

\subsection{Answer generation}

\begin{figure}[!h]
     \centering
     \includegraphics[width=0.9\linewidth]{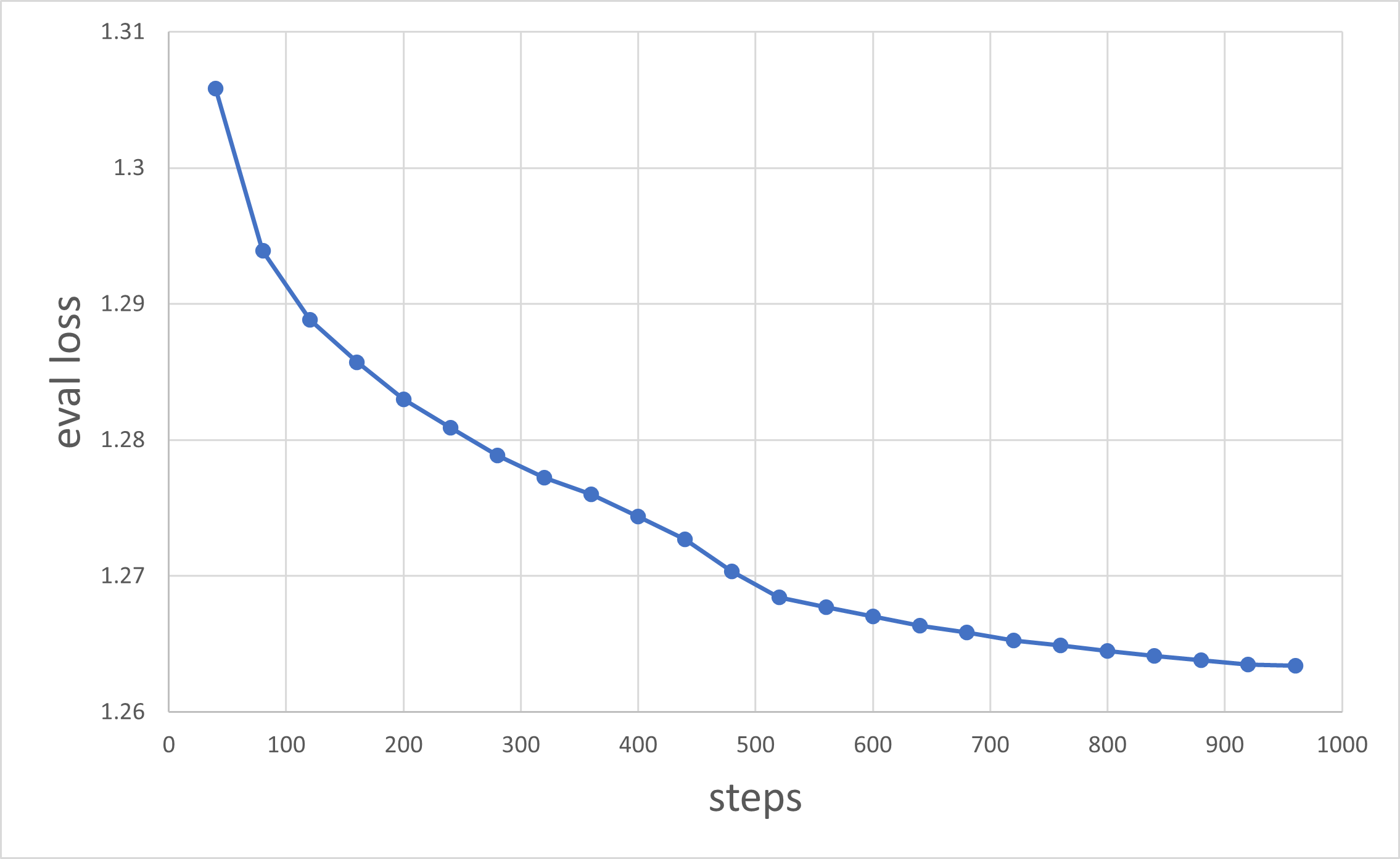}
     \caption{Evaluation loss for fine-tuning of Mistral 7B model on PubMedQA questions with generated and referenced answers}\label{Fig:EvalLoss}
\end{figure}

As we previously mentioned, we have fine-tuned Mistral 7B for question answering on questions coming from PubMedQA and answers generated using PubMed searches for relevant abstracts and GPT-3.5 for actual answer generation. The evaluation loss for the fine-tuning process can be seen in Figure \ref{Fig:EvalLoss}.

The fine-tuning of the Mistral 7B model improved the model's performance, making the generated answers comparable to those of much larger GPT-3.5 and GPT-4 models for the referenced question-answering task.

After manually comparing answers from GPT-3.5, GPT-4, and Mistral-7B-Instruct-v0.1-pqa to a test set of 50 questions and extracted abstracts, no model showed a clear advantage over the others. The quality, referenced abstracts, and length of the answers varied within each model and among the models. In terms of referenced abstracts, most of the time all three models referenced the same abstracts as relevant.

\subsection{Verification and hallucination detection}

The evaluation of the fine-tuned XLM-RoBERTa and DeBERTa model on the SciFact dataset that can be used for hallucination detection can be seen in Table \ref{EntailmentTable}. The model used 10\% of the data for validation and 10\% of the dataset for evaluation (test set). All three sets have homogenous distribution of the classes (36\%:42\%:22\% for NO\_EVIDENCE, SUPPORT and CONTRADICT classes respectively).

\begin{table}[!h]
\scriptsize
\centering
\caption{The evaluation of the entailment model fine-tuned from XLM-RoBERTa-large and DeBERTa-large model using SciFact dataset}

\begin{tabular}{|l|c|c|c|c|c|c|} 
\hline
& \multicolumn{3}{c|}{\textbf{XLM-RoBERTa}}& \multicolumn{3}{c|}{\textbf{DeBERTa}}  \\
& \textbf{Precision}& \textbf{Recall} & \textbf{F1-score} & \textbf{Precision}& \textbf{Recall} & \textbf{F1-score}  \\ \hline
\textbf{NO\_EVIDENCE}   & 0.91  & 0.96 & 0.95   & 0.88  & 0.86 & 0.87 \\\hline
\textbf{SUPPORT}   & 0.91  & 0.75 & 0.82   & 0.87  & 0.92 & 0.90  \\ \hline
\textbf{CONTRADICT}   & 0.59  & 0.81 & 0.68  & 0.88  & 0.81 & 0.85    \\ \hline
\textbf{Weighted Avg} & 0.87  & 0.85 & 0.85 & 0.88  & 0.88 & 0.88   \\ \hline
\end{tabular}
\label{EntailmentTable}
\end{table}

As can be seen from the table, the models exhibited state-of-the-art performance, surpassing the reported scores in \cite{wadden2020fact} for the label prediction task, and DeBERTa-large model showed superior performance compared to the RoBERTa-large. We use fine-tuned DeBERTa-large model for verification and hallucination detection. We also evaluated the SciFact label prediction task using the GPT-4 model, resulting in a precision of 0.81, recall of 0.80, and an F-1 score of 0.79. Therefore, our models outperformed GPT-4 model in zero-shot regime with carefully designed prompt for label prediction for the claims and abstracts in the SciFact dataset. It is important to note that the SciFact dataset contains challenging claim/abstract pairs, demanding a significant amount of reasoning for accurate labeling. Thus, in a real-use case where answers are generated by Mistral or another generative model, the task becomes easier. We believe that this model provides a good starting point for hallucination detection, as supported by our qualitative analysis of several pairs of generated claims and abstracts, which demonstrated good performance.

However, this model has some limitations. While it is capable of reasoning around negations, detecting contradicting claims, differing in just few words switching the context of the claim compared to the text of the abstract, proves to be a challenge. Additionally, we observe that neither model handle well situations where numerical values in claims are slightly different from the ones in the abstract.

\section{Conclusion}

In this short paper, we present the current progress on the \textbf{Verif.ai} project, an open-source generative search with referenced and verifiable answers based on PubMed. We describe our use of OpenSearch to create a hybrid search based on both semantic and lexical search methods, an answer generation method based on fine-tuning the Mistral 7B model, and our first hallucination detection and answer verification model based on fine-tuned DeBERTa-large model. However, there are still a number of challenges to be addressed and work to be done.

LLMs are rapidly developing, and performant, smaller LLMs, with larger context size are becoming more available. We aim to follow this development and use the best available open-source model for the task of referenced question-answering. We also aim to release early and collect user feedback. Based on this feedback, we aim to design an active learning method and incorporate user feedback into the iterative training process for both answer generation and answer verification and hallucination detection.

The model for hallucination detection and answer verification exhibits some limitations when it needs to deal with numerical values or perform complex reasoning and inference on abstracts. We believe that a single model may not be sufficient to verify the abstract well, but it may be the case that a solution based on a mixture of experts may be required \cite{jacobs1991adaptive,jiang2024mixtral}. To build user trust, we aim to offer several answer verification methods, some of which should be based on explainable AI and be easy for users to understand. In the future, this may include, for example, verification based on sentence similarity scores.

Currently, the system is designed for use in the biomedical domain and provides answers based on scientific articles indexed in PubMed. However, we believe that the system can be easily extended to other document formats and become a base for a personal, organizational, or corporate generative search engine with trustworthy answers. In the future, our version may incorporate additional sources, contributing to the trust and safety of the next generation internet.

\section{Availability}

Code created so far in this project is available on GitHub\footnote{\url{https://github.com/nikolamilosevic86/verif.ai}} under AGPLv3 license. Our fine-tuned qLoRA adapter model for referenced question answering based on Mistral 7B \footnote{\url{https://huggingface.co/filipealmeida/Mistral-7B-Instruct-v0.1-sharded}} is available on HuggingFace\footnote{\url{https://huggingface.co/BojanaBas/Mistral-7B-Instruct-v0.1-pqa}}. The verification models are available on HuggingFace\footnote{\url{https://huggingface.co/nikolamilosevic/SCIFACT_xlm_roberta_large}} \footnote{\url{https://huggingface.co/MilosKosRad/DeBERTa-v3-large-SciFact}}. More information on the project can be found on the project website: \url{https://verifai-project.com}.

\section*{Acknowledgment}
The project \textbf{Verif.ai} is a collaborative effort of Bayer A.G. and the Institute for Artificial Intelligence Research and Development of Serbia, funded within the framework of the NGI Search project under Horizon Europe grant agreement No 101069364.

\bibliographystyle{vancouver}
\bibliography{cas}

\begin{thebibliography}{10}

\bibitem{openai2023gpt4}
OpenAI. GPT-4 Technical Report; 2023.

\bibitem{jiang2023mistral}
Jiang AQ, Sablayrolles A, Mensch A, Bamford C, Chaplot DS, Casas Ddl, et~al.
\newblock Mistral 7B.
\newblock arXiv preprint arXiv:231006825. 2023.

\bibitem{bubeck2023sparks}
Bubeck S, Chandrasekaran V, Eldan R, Gehrke J, Horvitz E, Kamar E, et~al.
\newblock Sparks of artificial general intelligence: Early experiments with gpt-4.
\newblock arXiv preprint arXiv:230312712. 2023.

\bibitem{park2023generative}
Park JS, O'Brien J, Cai CJ, Morris MR, Liang P, Bernstein MS.
\newblock Generative agents: Interactive simulacra of human behavior.
\newblock In: Proceedings of the 36th Annual ACM Symposium on User Interface Software and Technology; 2023. p. 1-22.

\bibitem{touvron2023llama}
Touvron H, Martin L, Stone K, Albert P, Almahairi A, Babaei Y, et~al.
\newblock Llama 2: Open foundation and fine-tuned chat models.
\newblock arXiv preprint arXiv:230709288. 2023.

\bibitem{katz2023gpt}
Katz DM, Bommarito MJ, Gao S, Arredondo P.
\newblock Gpt-4 passes the bar exam.
\newblock Available at SSRN 4389233. 2023.

\bibitem{lewis2020retrieval}
Lewis P, Perez E, Piktus A, Petroni F, Karpukhin V, Goyal N, et~al.
\newblock Retrieval-augmented generation for knowledge-intensive nlp tasks.
\newblock Advances in Neural Information Processing Systems. 2020;33:9459-74.

\bibitem{wadden2020fact}
Wadden D, Lin S, Lo K, Wang LL, van Zuylen M, Cohan A, et~al.
\newblock Fact or fiction: Verifying scientific claims.
\newblock arXiv preprint arXiv:200414974. 2020.

\bibitem{huang2023survey}
Huang L, Yu W, Ma W, Zhong W, Feng Z, Wang H, et~al.
\newblock A survey on hallucination in large language models: Principles, taxonomy, challenges, and open questions.
\newblock arXiv preprint arXiv:231105232. 2023.

\bibitem{malaviya2023expertqa}
Malaviya C, Lee S, Chen S, Sieber E, Yatskar M, Roth D.
\newblock Expertqa: Expert-curated questions and attributed answers.
\newblock arXiv preprint arXiv:230907852. 2023.

\bibitem{boyko2023interdisciplinary}
Boyko J, Cohen J, Fox N, Veiga MH, Li JI, Liu J, et~al.
\newblock An Interdisciplinary Outlook on Large Language Models for Scientific Research.
\newblock arXiv preprint arXiv:231104929. 2023.

\bibitem{craswell2021ms}
Craswell N, Mitra B, Yilmaz E, Campos D, Lin J.
\newblock Ms marco: Benchmarking ranking models in the large-data regime.
\newblock In: Proceedings of the 44th International ACM SIGIR Conference on Research and Development in Information Retrieval; 2021. p. 1566-76.

\bibitem{jin2019pubmedqa}
Jin Q, Dhingra B, Liu Z, Cohen WW, Lu X.
\newblock Pubmedqa: A dataset for biomedical research question answering.
\newblock arXiv preprint arXiv:190906146. 2019.

\bibitem{dettmers2023qlora}
Dettmers T, Pagnoni A, Holtzman A, Zettlemoyer L.
\newblock Qlora: Efficient finetuning of quantized llms.
\newblock arXiv preprint arXiv:230514314. 2023.

\bibitem{jacobs1991adaptive}
Jacobs RA, Jordan MI, Nowlan SJ, Hinton GE.
\newblock Adaptive mixtures of local experts.
\newblock Neural computation. 1991;3(1):79-87.

\bibitem{jiang2024mixtral}
Jiang AQ, Sablayrolles A, Roux A, Mensch A, Savary B, Bamford C, et~al.
\newblock Mixtral of Experts.
\newblock arXiv preprint arXiv:240104088. 2024.

\end{thebibliography}
\vspace{12pt}

\end{document}